\documentclass[global,twocolumn,referee]{svjourarXiv}

\usepackage{graphics}

\begin{document}
\title{Cavity-enhanced direct frequency comb spectroscopy}

\author{Michael J. Thorpe and Jun Ye}

\institute{JILA, National Institute of Standards and Technology and
University of Colorado, Boulder, CO 80309-0440, USA\\ Department of
Physics, University of Colorado, Boulder, CO 80309-0390, USA\\}

\maketitle
\begin{abstract}
Cavity-enhanced direct frequency comb spectroscopy combines broad
spectral bandwidth, high spectral resolution, precise frequency
calibration, and ultrahigh detection sensitivity, all in one
experimental platform based on an optical frequency comb interacting
with a high-finesse optical cavity. Precise control of the optical
frequency comb allows highly efficient, coherent coupling of
individual comb components with corresponding resonant modes of the
high-finesse cavity. The long cavity lifetime dramatically enhances
the effective interaction between the light field and intracavity
matter, increasing the sensitivity for measurement of optical losses
by a factor that is on the order of the cavity finesse. The use of
low-dispersion mirrors permits almost the entire spectral bandwidth
of the frequency comb to be employed for detection, covering a range
of $\sim$10$\%$ of the actual optical frequency. The light
transmitted from the cavity is spectrally resolved to provide a
multitude of detection channels with spectral resolutions ranging
from a several gigahertz to hundreds of kilohertz. In this review we
will discuss the principle of cavity-enhanced direct frequency comb
spectroscopy and the various implementations of such systems. In
particular, we discuss several types of UV, optical, and IR
frequency comb sources and optical cavity designs that can be used
for specific spectroscopic applications. We present several
cavity-comb coupling methods to take advantage of the broad spectral
bandwidth and narrow spectral components of a frequency comb.
Finally, we present a series of experimental measurements on trace
gas detections, human breath analysis, and characterization of cold
molecular beams. These results demonstrate clearly that the wide
bandwidth and ultrasensitive nature of the femtosecond enhancement
cavity enables powerful real-time detection and identification of
many molecular species in a massively parallel fashion.
\end{abstract}
\section{Introduction}
Many applications ranging from scientific to commercial interests
will benefit from a highly sensitive, broad bandwidth, high spectral
resolution, and easy to use method for optical absorption
spectroscopy. In the commercial world alone, the number of
applications for optical detection of trace molecules has increased
enormously in the recent years, ranging from detection of
contaminants for semiconductor processing \cite{1} and human breath
analysis \cite{2} to air quality monitoring \cite{3} and detection
of biologically hazardous or explosive materials \cite{4}.

Each of these applications has been facilitated by the development
of new laser sources to access spectral domains of interest,
combined with cavity enhancement techniques for increased detection
sensitivity \cite{5}. For example, the advent of tunable laser
diodes has led to sensitive measurements of impurity levels in gases
such as phosphine used in semiconductor productions \cite{1} and for
monitoring ethylene in fruit storage facilities \cite{6}.
Subsequently, quantum cascade lasers and optical parametric
oscillators operating in the mid-infrared have been applied to human
breath analysis for health screening and optical detection of
explosives \cite{7}\cite{8}. These innovations represent remarkable
advances in our capability to monitor and understand our
environment. However, these technologies still operate in relatively
narrow spectral regions and thus the number of different molecules
that can be studied or detected with a single system is very small.
This severely limits our capability of performing surveyance on a
global scale in real time.

\indent In the scientific community, sensitive and high resolution
spectroscopy over a large spectral window is useful for many areas
of research. For instance, studies of atmospheric chemistry, climate
changes, and pollution can benefit a great deal from the capability
to monitor multiple molecules simultaneously \cite{9}\cite{10}.
Investigations of molecular potential energy surfaces, energy level
structure, local and eigen-mode dynamics, intramolecular energy
redistributions, as well as interaction dynamics and chemical
reactions all desire spectroscopic probes with high spectral
resolution and accuracy over a wide range of spectral coverage
\cite{11}\cite{12}. For the emerging field of ultracold molecules,
our ability to photoassociate cold molecules from pairs of free
atoms depends on extensive spectroscopic probes to find the most
efficient photoassociation pathways \cite{13}. To study a new class
of collision and reaction dynamics in the low energy limit
\cite{14}\cite{15}, the need for higher resolution and sensitivity
while under a broad spectral coverage is now of more urgent
importance \cite{16}\cite{17}.

\indent Optical frequency combs based on mode-locked femtosecond
lasers provide hundreds of thousands of sharp spectral components -
frequency comb teeth - evenly distributed across hundreds of
nanometers of optical spectral bandwidth. Direct frequency comb
spectroscopy takes advantage of this unique property of frequency
comb to recover high resolution spectral features across the entire
manifold of atomic and molecular energy levels of interest
\cite{18}\cite{19}\cite{20}. The development of mode-locked fiber
lasers has led to robust optical frequency comb sources capable of
continuous operations without user intervention \cite{21}. Frequency
comb-based systems can also employ high peak intensities available
from the pulsed output to enable efficient nonlinear frequency
conversions for easy access to spectral regions spanning from the UV
to the far infrared \cite{22}\cite{23}. While techniques such as
fourier transform infrared spectroscopy (FTIR) also offer broad
spectral coverage with reasonably good spectral resolution, the
intrinsically parallel detection capability enabled by a frequency
comb brings a truly revolutionary impact to the efficiency of signal
collection. Recent approaches to molecular detection that use broad
bandwidth frequency combs have indeed created parallel detection
schemes that in a single shot record large spectral bandwidths
\cite{24}\cite{25}. The only missing piece in this powerful
combination for spectroscopy is high sensitivity, which can be
satisfactorily addressed via the use of high-finesse, broadband,
low-dispersion enhancement cavities \cite{26}\cite{27}.

Frequency comb based spectroscopic systems have made significant
progress recently \cite{28}\cite{29}\cite{30}. While they are not
yet as simple as an FTIR-like spectrometer, the demonstrated
capability of providing cavity-enhanced sensitivity over hundreds of
nanometers of optical wavelength has stimulated a great interest in
the spectroscopy community. It is our hope that this review article
will provide useful material to facilitate the spread of this new
spectroscopic technique to the wider community. In the article we
present a review of cavity-enhanced direct frequency comb
spectroscopy (CE-DFCS), discussing the principles of operations
followed by several examples of applications. To provide useful
technical information, a number of optical frequency comb sources
will be surveyed, including titanium sapphire lasers, erbium fiber
lasers, yetterbium fiber lasers, and other comb sources generated
through nonlinear frequency conversion. Also, we will present a
number of methods for coupling a frequency comb to an optical cavity
and discuss various techniques to provide parallel detections of
frequency-resolved absorption information transmitted from the
optical cavity. Methods for increasing the spectral bandwidth and
reducing acquisition time in future systems will be discussed.
Finally, a series of CE-DFCS measurements will be presented to
demonstrate the current capabilities of these systems.

\section{Frequency comb properties and sources}
\subsection{Comb properties}
Many of the same properties that make the frequency comb an
excellent tool for counting optical frequencies and transferring
coherence across vast spectral regions also make it an ideal light
source for broadband, cavity enhanced spectroscopy.  Frequency combs
are capable of covering very large spectral ranges of hundreds of
terahertz, thus enabling investigations of global energy level
structure for multiple species. At the same time, the spectrally
sharp and precisely defined frequency structure of the comb
components allow high-resolution and frequency-calibrated
spectroscopy across massively parallel detection channels. For the
same reason, the frequency comb is ideal for sensitive intracavity
spectroscopy since it can be efficiently coupled to a corresponding
set of resonant modes of a high finesse optical cavity across a wide
spectral region.

\begin{figure}[h]
\resizebox{0.5\textwidth}{!}{%
 \includegraphics{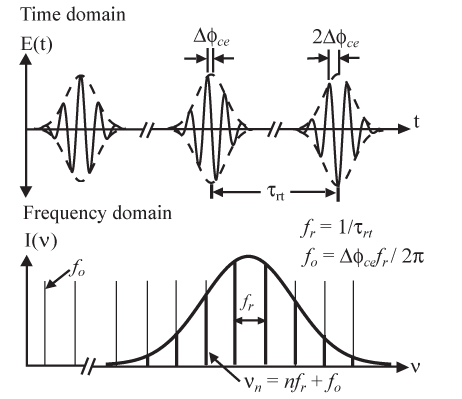}
} \caption{The time and frequency domain properties of an optical
frequency comb.  The top panel shows the time domain representation
of a mode-locked pulse train including the temporal spacing between
pulses ($\tau_{rt}$) and the evolution of the carrier envelope phase
($\Delta \phi_{ce}$).  The bottom panel shows the frequency domain
picture where the evenly spaced comb modes with frequencies $\nu_n$
lie beneath a gaussian spectral envelope.}
\end{figure}

The frequency of each comb component is defined by the repetition
frequency ($f_r$) and the carrier-envelope offset frequency ($f_o$)
via the simple relation $\nu_n$ = n$f_r$ + $f_o$ \cite{31}. The time
and frequency domain representations of a comb are illustrated in
Fig. 1. For optical frequencies in the visible and near IR, the comb
mode index $n$ takes on values ranging from $10^5$ to $10^7$,
depending on $f_r$. The typical range for $f_r$ is 10 MHz $<$ $f_r$
$<$ 10 GHz while $f_o$ can take on values of
-$f_r/2$$<$$f_o$$<$$f_r/2$. Precise control of these two degrees of
freedom establishes the comb as an accurate frequency tool for
atomic and molecular spectroscopy and also permits efficient
coupling of the comb to the modes of a high finesse passive optical
cavity. To implement control, one first needs to measure $f_r$ and
$f_o$. $f_r$ can be determined directly from the pulse train
incident on a fast photodiode while $f_o$ requires nonlinear
interferometry \cite{31} \cite{32}. Two stable radio frequency
references can be used subsequently to stabilize $f_r$ and $f_o$ via
servo transducers acting on the laser cavity length and the
intra-cavity dispersion. A frequency comb can also be stabilized
directly to a passive optical cavity by detecting the cavity
response to the incident comb spectrum in the optical domain
\cite{33}, as discussed in later sections of this paper.

\subsection{Comb sources and their applications to spectroscopy}
Many different types of mode-locked laser sources exist to provide
coverage for a wide variety of spectral regions from blue to the mid
IR. When nonlinear frequency conversion and spectral broadening
techniques are implemented, the spectral coverage can be extended
well into the VUV and the far infrared. The diversity of spectral
regions that can be accessed allow frequency comb spectroscopy to be
performed on a large variety of atomic and molecular systems. For
the purposes of this review, we will restrict discussions to
titanium sapphire lasers, erbium fiber lasers, and yetterbium fiber
lasers. Nonlinear frequency conversion to other spectral regions
based on these laser sources will also be briefly discussed. Figure
2 shows spectral overlaps between various frequency comb sources and
relevant regions for useful atomic and molecular spectroscopy.

\begin{figure}[h]
\resizebox{0.5\textwidth}{!}{%
 \includegraphics{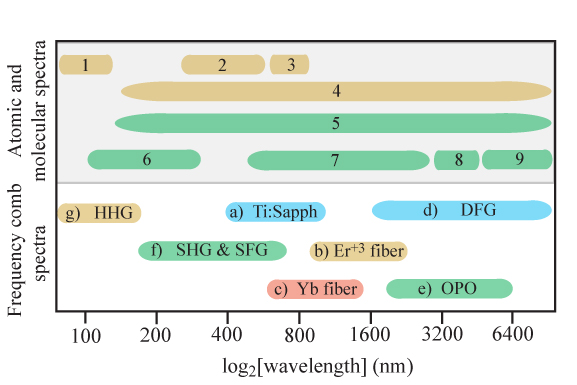}
 } \caption{Spectral coverage for several types of mode-locked
lasers and corresponding atomic and molecular transitions to which
they can be applied. 1, transitions from ground states of noble gas;
2 and 3, transitions from ground states of alkaline earth or alkali
atoms; 4 and 5, transitions between excited atomic or molecular
states; 6, transitions from ground states of molecules; 7, overtones
of molecular ro-vibrational transitions; 8 and 9, fundamental
vibrations of light-nuclei or heavy-nuclei molecules. a) Ti:sapphire
mode-locked lasers, after nonlinear spectral broadening, covers the
entire visible range from 400 to 1200 nm; b)Erbium-doped fiber
lasers (centered around 1550 nm) can be made to cover 1000 - 2400 nm
via spectral broadening; c)Ytterbium-doped fiber lasers (centered
around 1030 nm) can cover 600 - 1600 nm via spectral broadening.
Each of these lasers can have their spectral coverage further
extended via nonlinear frequency conversion using techniques listed
in (d), (e), (f), and (g). We thus have the possibility of
performing comb spectroscopy for nearly all of the interested atomic
and molecular transitions.}
\end{figure}

\indent While the gain bandwidth of Ti:sapphire lasers cover only
700-950 nm, subsequent super-continuum generations in photonic
crystal and tapered telecommunications fibers
\cite{23}\cite{34}\cite{35}, and the later development of
octave-spanning Ti:sapphire lasers via intracavity nonlinear phase
modulation allowed spectral coverage extending from the deep blue to
the near Infrared (NIR) \cite{36}\cite{37}\cite{38}.  As the first
generation of precisely controlled optical frequency combs were
based on Ti:sapphire lasers, it is natural that they played
important roles in the first demonstrations of direct frequency comb
spectroscopy \cite{18}. Frequency comb based cavity-enhanced
absorption spectroscopy (CEAS) and cavity ring-down spectroscopy
(CRDS) were also first performed with Ti:sapphire combs on molecular
overtone transitions \cite{26}\cite{27}. The second-harmonic of a
Ti:sapphire comb was scanned over a low-finesse cavity to measure
high overtones of C$_2$H$_2$ in the blue visible region \cite{39}.
Frequency down conversion of Ti:sapphire combs has been used to
produce infrared frequency combs via difference frequency generation
(DFG) and optical parametric oscillation (OPO) techniques \cite{40}
\cite{41} \cite{42}. Such infrared combs have been used to probe the
strong fundamental vibration transition in molecules \cite{43}.
Finally, high harmonic generation (HHG) has been used to generate
frequency combs in the vacuum ultra-violet regions
\cite{44}\cite{45}, and with improvement in the power \cite{46} they
could be used to probe ground state transitions of noble gases
\cite{47}.

\indent  More recently, mode-locked erbium and ytterbium fiber
lasers have been developed that generate highly stable frequency
combs which are pumped by inexpensive laser diodes \cite{48}
\cite{49} \cite{50} \cite{51} \cite{52}. These lasers can
conveniently produce super-continuum spectra via coupling to highly
nonlinear fibers. They have been frequency converted to cover the
spectral domain extending from the infrared to ultra-violet, using
the same techniques described for Ti:sapphire \cite{53}\cite{54}.
The fiber lasers are compact, reliable, and robust, with turn-key
operations. Thus they have a very promising future for commercial
applications. Recent work has used an erbium fiber comb to perform
CE-DFCS on low molecular overtones for trace detection of a wide
variety of molecules \cite{28}\cite{30}. Finally, fiber combs are
conveniently located in the 1.0 - 1.5 $\mu$m spectral domain,
overlapping with strong first-order vibration overtones of many
molecules. They are also ideal for frequency conversion to extend
the coverage of frequency combs to the mid and far infrared,
important regions to cover fundamental vibrational transitions of
molecules.

\section{Useful properties for passive optical cavities}
Important properties for passive optical cavities used in CE-DFCS
include the finesse for detection sensitivity, the intra-cavity
dispersion for comb-cavity coupling, and the spectral bandwidth for
useable spectroscopy windows. In this section we analyze these
properties for a cavity constructed from two highly reflective
mirrors with quarter-wave stack coatings. This analysis will
illustrate why high finesse, low dispersion, and large spectral
bandwidth are desirable qualities for a CE-DFCS cavity. We will
discuss their intrinsic relationships and necessary compromises we
need to make to achieve optimum detection. Another form of a
high-finesse, low-dispersion optical cavity, i.e., a
retro-reflecting prism cavity \cite{55}, will be mentioned briefly
at the end of this section.

\subsection{Finesse}
The most important property of a cavity is its finesse. The finesse
determines the enhancement factor for the intracavity absorption
signal that can be recovered from the cavity transmission.
Therefore, precise knowledge of the cavity finesse is important for
making accurate measurements of the intracavity absorption. The
cavity finesse $\mathcal{F}$ is proportional to the inverse of the
total intracavity loss. For a cavity constructed from two mirrors
with negligible scattering losses, the cavity finesse is related to
the mirror reflectivities R$_1$ and R$_2$ by the familiar equation
$\mathcal{F} = \pi \sqrt[4]{R_{1}R_{2}}/(1-\sqrt{R_{1}R_{2}})$
\cite{56}.

\indent For low-loss cavities ($\mathcal{F}>$ 1000), the cavity
finesse is usually measured via CRDS \cite{57}. CRDS measurements
are performed by injecting laser light into a resonant mode of the
cavity and then rapidly switching off the incident light. Without
input, the intracavity power gradually decays at a rate that is
determined by the ratio of the total cavity loss and the cavity
round-trip time, i.e., the linewidth of a cavity resonance mode. The
cavity decay time $\tau _{cavity}$ and finesse $\mathcal{F}$ are
thus related by $\mathcal{F}=2\pi \tau _{cavity}$ $FSR$, where $FSR$
represents the cavity free-spectral-range frequency.

It is straightforward to use a single-frequency CW laser for CRDS
measurements. For fast switching of the light source, the laser
frequency can be swept quickly across the cavity resonance such that
the time the laser spends on resonance with the cavity is shorter
than the cavity lifetime. Another way of switching the incident
light is to use an acousto-optic or electro-optic modulator that can
provide switching times of less than a few hundred nanoseconds,
sufficiently shorter than a typical high-finesse cavity lifetime.
CRDS measurements can also be performed with an optical frequency
comb if the mode spacings of the cavity and the comb are roughly
matched. The frequency comb components can then be frequency-scanned
across the corresponding cavity modes just as in the case for a CW
laser. When the cavity transmitted light is frequency resolved using
a grating or some other forms of dispersion, the information about
the cavity finesse, and thus the intracavity loss, is obtained
simultaneously across the entire spectrum of the comb \cite{27}.

\begin{figure}[h]
\resizebox{0.5\textwidth}{!}{%
 \includegraphics{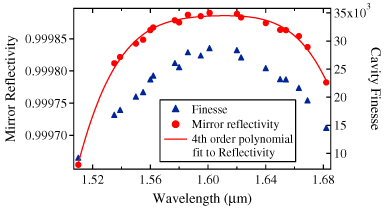} }
 \caption{The finesse, and the corresponding cavity mirror
reflectivity, of an optical enhancement cavity spanning from 1.51 to
1.68 $\mu$m measured by a mode-locked Er$^{+3}$ fiber laser using
frequency-resolved cavity ring-down spectroscopy.}
\end{figure}

A set of measurements using fast sweeping of the comb frequencies
and a grating disperser after the cavity are shown in Fig. 3. Here,
a frequency comb from a mode-locked Er-fiber laser was used to
couple into a cavity constructed from two mirrors with peak
reflectivity at 1.6 $\mu$m \cite{29}. A grating was used in cavity
transmission to provide a 1-nm spectral resolution. A beam splitter
was used to direct a portion of the cavity transmission to an
optical spectrum analyzer for wavelength calibration of the
ring-down measurements.

\subsection{Cavity mode frequencies and dispersion}
The intracavity dispersion of an optical cavity is important,
especially for frequency comb-based spectroscopy, as it determines
the spectral bandwidth over which the cavity modes and the comb
components can be overlapped simultaneously.  While the comb modes
are evenly spaced in the frequency domain, $\nu_{n}=nf_{r}+f_{o}$,
the cavity mode spacing is frequency dependent,
$FSR=c/(2L+c(\partial\phi/\partial\omega)|_{\omega})$, where $L$ is
the cavity length and $c$ is the speed of light. The degree to which
the cavity $FSR$ is wavelength (or optical frequency)-dependent is
determined by the intracavity dispersion term
$c(\partial\phi/\partial\omega)|_{\omega}$
\cite{58}\cite{59}\cite{60}. Clearly, low intracavity dispersion is
important for coupling a broad-bandwidth comb spectrum to respective
cavity modes. Furthermore, precise characterizations of the
intracavity dispersion are essential for the cavity filtered
detection scheme described in Section 6.3. For now we describe a
method of using the frequency comb itself to precisely measure the
intracavity dispersion. We will discuss the application of these
measurements in later sections.

\begin{figure}[h]
\resizebox{0.5\textwidth}{!}{%
 \includegraphics{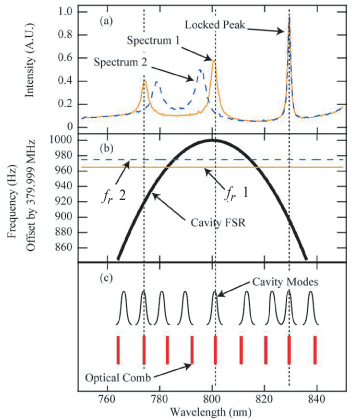}
} \caption{A technique for using the frequency comb for measuring
intracavity dispersion. (a)  The cavity transmission spectrum for
two values of $f_r$, showing the spectral regions where the comb
modes and cavity modes overlap.  While comb/cavity mode overlap is
forced at the locking wavelength near 830 nm, the other cavity
transmission peaks arise from the frequency dependence of the cavity
$FSR$. (b) The frequency dependent cavity $FSR$ and values of $f_r$
that give rise to the transmission spectra in (a).  (c) A schematic
of the cavity and comb modes for the case of Spectrum 1 in (a),
illustrating that cavity transmission peaks occur only where there
is overlap between the cavity and comb modes.}
\end{figure}

\indent A schematic of the comb-based cavity dispersion measurement
is shown in Fig. 4.  The comb acts as a frequency ruler to determine
the cavity mode structure and the wavelength-dependent $FSR$. First,
$f_r$ of the laser is detuned away from the cavity $FSR$ at 800 nm
as shown in Fig. 4 b). A frequency discrimination signal between the
cavity modes and frequency combs is recovered and serves to lock the
comb to the cavity near the edge of the cavity/comb spectrum (830 nm
in this case). This locking wavelength becomes a reference from
which the cavity $FSR$ is compared to the laser $f_r$. As one
gradually moves away from 830 nm to other parts of the comb
spectrum, the corresponding comb components will spectrally walk
away from their respective cavity resonance modes. However, after a
sufficiently large spectral gap, a set of comb lines will again come
on resonance with cavity modes. Overlaps between the comb
frequencies and cavity modes that lead to cavity transmissions are
schematically illustrated in Fig. 4 c). A spectrometer is used to
measure the wavelength-resolved intensity of the light transmitted
from the cavity, as shown in Fig. 4 a) for two different values of
$f_r$.

The frequency separation between the $m^{th}$ cavity mode and the
$m^{th}$ comb component from the locking point (830 nm) can be
determined by the walk-off integral \cite{60},
\begin{equation}
\Delta \nu_m =\int_{\nu_l}^{\nu_m}\bigg(\frac {FSR(\nu)}{f_r(\nu_m)
}-1\bigg) d\nu .
\end{equation}
Here $FSR$($\nu$) is the wavelength (frequency)-dependent cavity
$FSR$ that is to be determined. $\Delta \nu_m$ is the frequency gap
between the m$^{th}$ comb component and the m$^{th}$ cavity mode
from the locking wavelength. $\nu_l$ is the frequency of the locking
wavelength and $\nu_m$ is the frequency of the m$^{th}$ comb mode.
When $\Delta \nu_m$ = $kf_r$ ($k$ = ..., -2, -1, 0, 1, 2, ...), the
$m^{th}$ comb frequency overlaps with the $(m+k)^{th}$ cavity mode.
This comb component then appears as a peak in the cavity transmitted
spectrum. With the laser locked to the cavity at 830 nm, $f_r$ is
scanned and the wavelength values of the cavity transmission peaks
are recorded for each value of $f_r$. By differentiating the
walk-off integral with respect to the comb mode number we obtain
\begin{equation}
FSR(\nu_t)=f_r(\nu_t)+(\nu_t -\nu_l)\frac{d}{d\nu_t}f_r(\nu_t).
\end{equation}
Here, the values of $f_r$ that give rise to cavity transmission
peaks at frequencies $\nu_t$ are used to determine the frequency
dependance of the cavity $FSR$. Finally, a simple equation
\begin{equation}
GDD(\nu)=\frac{d}{d\nu}\frac{1}{FSR(\nu)}
\end{equation}
is used to calculate the intracavity group-delay-dispersion from the
frequency dependent cavity $FSR$.  Fig. 5 shows a measurement of the
intracavity dispersion for an evacuated two-mirror cavity centered
at 800 nm using the comb technique.
\begin{figure}[h]
\resizebox{0.5\textwidth}{!}{%
 \includegraphics{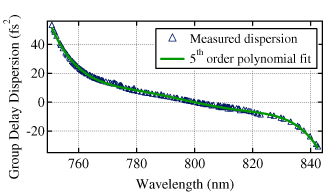}
} \caption{Measurement of the dispersion of a two-mirror cavity with
peak reflectivity at 800 nm, using the comb-based cavity dispersion
measurement technique.}
\end{figure}

The precision of this dispersion measurement technique is directly
related to the cavity finesse. Higher finesse leads to more sharply
peaked features in cavity transmission and reduced uncertainties in
the determination of the center wavelength of the transmission peak.
In the example for Fig. 5, the cavity under measurement had a
finesse of 3000 and the achieved experimental precision for
intracavity dispersion measurement was better than 1 fs$^2$.

\subsection{Spectral bandwidth}
\indent The spectral bandwidth of a cavity refers to a spectral
window in which the cavity has a high finesse and nearly uniform
cavity $FSR$ that are important for cavity-enhanced frequency comb
spectroscopy. The spectral bandwidth provided by quarter-wave stack
Bragg reflectors depends on the refractive index contrast of the two
materials that comprise the stack.  Typical bandwidths are around
$\sim$15\% of the center optical frequency where the mirrors have
their peak reflectivity. However, the spectral coverage of modern
optical frequency combs can be very large ($\sim$90\% of the central
frequency).  As a result, optical cavity designs currently available
limit the spectral coverage that can be achieved with CE-DFCS.\\

In response to this, efforts toward improved designs for low-loss
and low-dispersion mirrors have been intense over the past few years
\cite{61}.  Also, new designs for high-finesse cavities that use
prism retroreflectors instead of mirrors are under development.
Since prism retroreflectors are based on broad bandwidth effects,
such as Brewsters angle and total internal reflection, these
cavities could provide spectral bandwidths of 80\% of the center
optical frequency with a finesse of up to 50,000 \cite{55}. Such
cavities have increased dispersion since a portion of the cavity
mode is inside of the prism.  However, careful design of the cavity
prisms, including minimizing the prism size and the use of low
dispersion materials, looks promising for providing simultaneous low
dispersion and high finesse over large spectral bandwidths.

\section{Cavity/comb coupling}
In this section we discuss techniques that are used to couple the
frequencies of an optical comb to the resonant modes of an optical
cavity. Three techniques will be discussed, including (1) lock of
the comb components to the cavity modes, (2) a low amplitude, fast
sweep of the comb components over the cavity modes, and (3)
precision scan between initially mismatched $FSR$ and $f_r$ such
that one comb component leaks out of the cavity at a time. Each of
these techniques offers different compromises between benefits and
limitations for performing CE-DFCS. Practical applications will
dictate the choice for the right approach.

\subsection{Comb locked to cavity}
 The most intuitive method for coupling a frequency comb to an
 optical cavity is a precise and simultaneous overlap of all of the comb
 frequencies with corresponding cavity modes (Fig. 6).  To perform this
 type of coupling requires that the optical cavity maintains the same optical path
 length as that of the frequency comb laser (i.e. $FSR$ = $f_r$).  Also, the
 dispersion of the optical cavity must be sufficiently low to ensure a large number of comb
 modes overlap with their corresponding cavity modes.

\begin{figure}[h]
\resizebox{0.5\textwidth}{!}{%
 \includegraphics{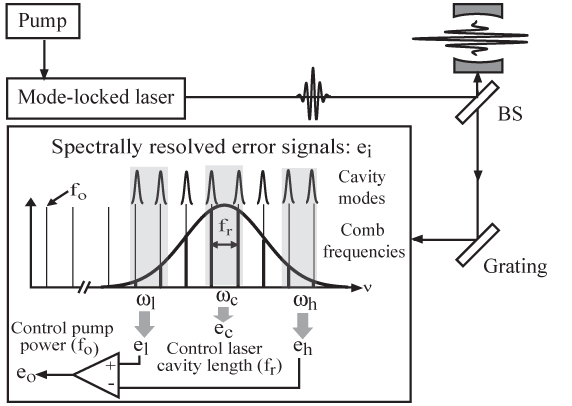}
} \caption{A frequency comb can be locked to an optical cavity by
generating spectrally resolved cavity-locking error signals.  An
error signal at the center frequency is used to lock a collection of
comb modes at $\omega_c$ to be on resonance with their respective
cavity modes. Error signals at the high ($\omega_h$) and low
($\omega_l$) ends of the comb spectrum can be used to maintain a
correct value of $f_o$. Adjustment of $f_o$ and $f_r$ (via the
optical lock at $\omega_c$) ensures an optimal overlap of the comb
and cavity modes over a wide spectral bandwidth.}
\end{figure}

For a broadband frequency comb, two servo loops are needed to lock
the comb to a cavity \cite{32}\cite{62}, one for $f_r$ and the other
for $f_o$. The two servo error signals are generated through
spectrally resolved detection of the cavity-comb interaction as
shown in Fig. 6. For robust lock, we use a modulation/lock-in
detection approach. For example, an EOM can be used to place
modulation sidebands on the frequency comb, or the cavity length can
be dithered rapidly.  A collection of comb modes at the center of
the spectrum is detected, after their interaction with their
respective cavity modes. This generates an error signal that locks
the optical comb frequencies to the cavity modes by acting on $f_r$.
The interaction of the high and low frequency ends of the comb
spectrum with the cavity can be used to lock f$_o$ to the value that
results in the maximum spectral bandwidth coupled into the optical
cavity. These two loops act together to optimize the coupling
between the comb and cavity.

\indent  The main benefit of locking the comb to the cavity is that
the optical power transmitted through the cavity is nearly that of
the beam incident to the cavity. Furthermore, the cavity
transmission is continuously present. These two features yield a
high signal to noise ratio ($S/N$) for measurements of the cavity
transmitted beam and a fast averaging time due to the continuous
presence of the cavity transmission. The largest drawback is that
the spectral bandwidth of the comb that is coupled into the cavity
is typically limited to $<10\%$ of the center frequency. Also, the
continuous coupling of the comb to the cavity is very sensitive to
acoustic and vibration-induced frequency noise that gets converted
to a substantial amount of intensity noise on the cavity
transmission if the servo loops do not have sufficient gains.

\subsection{Comb swept around cavity}

The second method for coupling comb frequencies to cavity modes
involves a periodic injection of the comb into the cavity.  This is
accomplished by introducing a low amplitude, quick frequency sweep
between the corresponding modes of the cavity and the comb
\cite{26}\cite{27}. This type of coupling has several advantages and
some drawbacks compared to directly locking the comb to the cavity.
The first advantage is that the requirement on the intracavity
dispersion is relaxed. Since the comb and cavity modes are being
swept over each other, it is no longer required that all of the comb
and cavity modes are exactly overlapped simultaneously. Instead,
different modes are allowed to come on resonance at slightly
different times, thus increasing the useful spectral bandwidth that
is recovered in cavity transmission. With a reasonably low
dispersion, high finesse, and fast frequency sweeping, the time it
takes for different comb modes to come on resonance with the cavity
can be shorter than the cavity lifetime.

In the case of Fig. 7, 100 nm of comb spectrum centered at 1.55
$\mu$m is swept over the cavity modes.  Due to low intracavity
dispersion and optimal matching of the comb to the cavity, all comb
frequencies come onto resonance with the cavity at nearly the same
time. Under these conditions, the cavity transmission signal is
sharply peaked. The amplitude of the frequency sweep can be made
very small such that the comb is on resonance with the cavity for a
large fraction of the sweep time. By using the smallest possible
sweep amplitude, the power contained in the cavity transmission is
maximized.

\begin{figure}[h]
\resizebox{0.5\textwidth}{!}{%
 \includegraphics{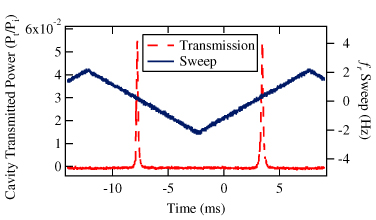}
} \caption{Cavity transmission for a 100 nm input spectrum coupled
to a $\mathcal{F}$ = 30000 cavity.  Due to low dispersion and
optimal cavity/comb matching, the entire 100 nm incident spectrum
becomes resonant with the cavity at nearly the same time during a
frequency sweep.}
\end{figure}

Another advantage of sweeping over the resonance is that it largely
avoids intensity noise on the cavity transmitted beam. Under the
locked condition, mechanical vibration-induced cavity length
fluctuations cause amplitude noise on cavity transmission, which can
place a limit on the detection $S/N$ of the transmitted beam. By
sweeping at rates larger than the Fourier frequencies of dominant
mechanical noise processes, the intensity noise of cavity
transmission is drastically reduced.

The final advantage of fast sweeping is the ease of control to
maintain the desired coupling between comb and cavity modes. This
approach only has to stabilize the time at which the comb modes come
onto resonance with the cavity modes during the sweep cycle. As a
result, a low bandwidth servo acting on the cavity or laser length
is sufficient for maintaining a constant sweep-based comb-cavity
coupling. An example of this type of servo is given in Fig. 8. The
cavity transmission is compared against a preset DC threshold in a
comparator, generating a train of pulses that indicate when the comb
is resonant with the cavity. These pulses become the 'clock' input
for a subsequent D-flip flop. The 'D' input of the flip flop is
provided by a TTL synchronization signal for the sawtooth sweep
waveform. The output is high when sawtooth is increasing in
amplitude and low when it is decreasing. The output of the flip flop
can then be integrated to indicate the separation between successive
cavity transmission peaks.  A DC offset voltage can then be used to
lock the separation of successive cavity transmission peaks to any
desired value.  The value of $f_o$ is measured either by the
interaction of the comb with the cavity, or by $f$ - $2f$
interferometry and is then set to a value that optimizes the
coupling of the comb to the cavity.
\begin{figure}[h]
\resizebox{0.5\textwidth}{!}{%
 \includegraphics{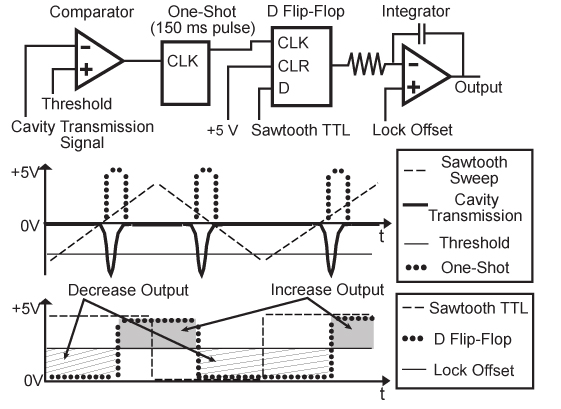}
} \caption{A schematic of the servo used to maintain the position at
which the comb and the cavity come onto resonance during each sweep
cycle.}
\end{figure}

\indent The primary disadvantage of sweep-based coupling is the loss
of duty cycle for useful cavity transmission. Making a faster sweep
reduces the intensity noise, but the power transmitted through the
cavity is also reduced. Therefore it is necessary to choose an
optimum sweep rate that maximizes the overall $S/N$ of the cavity
transmission by making a compromise between the intensity noise and
the transmission power. For systems implemented in our lab, a sweep
rate of 1-2 kHz is optimal such that the cavity is allowed to
build-up intensity for 1/10$\tau_{cavity}$. Of course, the optimal
settings depend on the bandwidth of the sweep actuator, and the
degree of isolation of the cavity and laser system from mechanical
vibrations of their environment. Less mechanical noise allows for a
longer intracavity build-up time and higher optical power in the
cavity transmitted beam.

\subsection{Precision sweep of laser $f_{r}$}

The final method of coupling the comb to cavity involves purposely
mismatching the cavity $FSR$ and the laser $f_r$.  The comb modes
are then swept precisely through successive cavity modes such that
the cavity acts as a frequency filter in transmission, allowing one
or a handful of comb components to be transmitted at a time. This
allows serial detection of individual frequency comb components
across the comb spectrum \cite{28}.

\begin{figure}[h]
\resizebox{0.45\textwidth}{!}{%
 \includegraphics{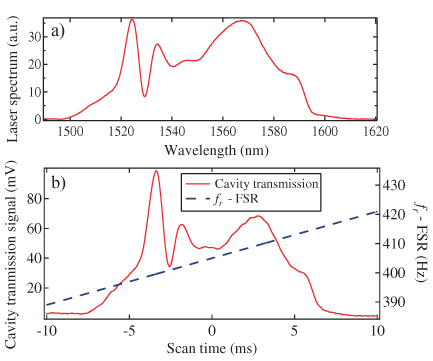} }
\caption{a) The optical spectrum of a mode-locked Er$^{+3}$ fiber
laser measured with an optical spectrum analyzer. b) The cavity
transmission signal as a function of the laser $f_r$ detuning form
the cavity $FSR$. The resulting cavity transmission signal recorded
over time is not sharply peaked, but rather resembles the original
spectrum of the mode-locked laser source.}
\end{figure}

To illustrate this approach, we demonstrate the measurement of the
spectrum of our mode-locked fiber laser using two distinctly
different methods (Fig. 9). Fig. 9a displays the spectrum of the
mode-locked laser recorded by a traditional optical spectrum
analyzer. In comparison, Fig. 9b shows the cavity transmission
signal for a detuning between $FSR$ and $f_r$ scanned in the range
of 390 to 420 Hz. This scan of the laser $f_r$ lasted about 20 ms.
The resulting cavity transmission signal recorded over this time
interval closely resembles the optical spectrum of the mode-locked
laser. The optical frequency resolution and spectral bandwidth
provided by the cavity filtering scheme is determined by the laser
$f_r$, the detuning of the cavity $FSR$ and $f_r$ ($\delta f$ =
$f_r$ - $FSR$), and the cavity linewidth ($\Delta \nu$). The
resolution is given by $\Delta \nu$ for $\Delta \nu < \delta f$; or
($\Delta \nu/\delta f$)$f_r$ for $\Delta \nu
> \delta f$. The first scenario corresponds to the case where only
one comb mode is resonant with the cavity at a time while the second
case gives the resolution when several adjacent comb modes are
simultaneously resonant with the cavity.\\

The spectral bandwidth that can be achieved in a single scan is
determined by $\delta f$ and $f_r$.  Figure 10 illustrates that if
the comb is on resonance with the cavity at a particular frequency,
$\delta f$ will initially cause the comb modes to become detuned
from the cavity modes in the neighboring spectral region. However,
after a frequency interval of $(f_r/\delta f) f_r$ = $f_r^2/\delta
f$ the comb will once again become resonant with the cavity modes.
This relation can be easily derived by noting that $\Delta N f_r$ =
$(\Delta N+1)FSR$ = $(\Delta N+1)(f_r - \delta f)$, hence the number
of modes contained within the spectral bandwidth is $\Delta N$
$\approx$$f_r/\delta f$.

When more than one frequency is transmitted from the cavity at a
time, there is no longer a unique mapping between scan time and the
frequency in cavity transmission. To avoid ambiguity in such a case,
a dispersive element with a coarse resolution (only need to be
better than $f_r^2/\delta f$) should be placed between the cavity
transmission and the detector. To obtain a precise calibration of
the cavity transmitted wavelength during a scan of $f_r$ requires
that the frequencies of the cavity modes be known to high precision.
For this, the dispersion measurement technique described in Section
3.2 can be used to provide a high precision measurement of the
wavelength dependence of the cavity mode frequencies.  Figure 10
illustrates the frequency domain picture of cavity filtering of a
comb where $f_r$ = 100 MHz, $\Delta\nu=3$ kHz, and $\delta f$ = 1
kHz.

\begin{figure}[h]
\resizebox{0.5\textwidth}{!}{%
 \includegraphics{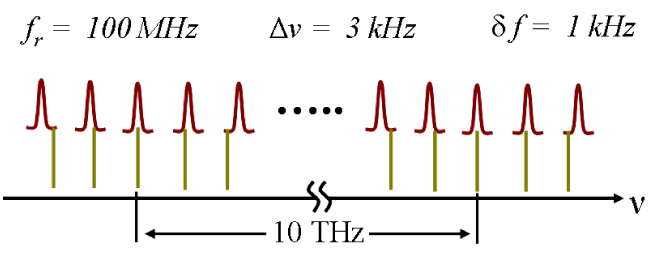} }
\caption{ An example of the resolution (300 MHz) and spectral
bandwidth ($f_r^2/\delta f$ = 10 THz) provided by cavity filtering
for $f_r$ = 100 MHz, $\delta f$ = 1 kHz, and $\Delta \nu$ = 3 kHz.}
\end{figure}

To achieve the actual cavity-linewidth-limited spectral resolution
for $\Delta \nu < \delta f$, the frequency difference between the
cavity and comb modes needs to be stabilized to within the cavity
linewidth. If the jitter of the cavity or comb modes are larger than
the cavity linewidth, then the value of ($\Delta \nu$) used for
determining the resolution should be replaced with the frequency
jitter. Since cavities used for sensitive detection can have
linewidths in the kHz range, stabilizing the jitter below this level
is straightforward in a research laboratory but could be challenging
in field applications.

Compared to the frequency sweep technique for coupling the comb to
the cavity, the approach of precision $f_r$ scan typically uses
slower sweep rates of the comb modes.  It is therefore more
susceptible to cavity/comb coupling noise arising from mechanical
vibrations. Furthermore, this approach takes measurements in a
serial manner, prolonging the measurement time. However, this
technique does have the benefit of spectrally resolving the cavity
transmitted signal without the need for an external spectrometer, a
unique feature not shared by the two other coupling techniques.

\section{Measurement schemes}
\label{sec:1} This Section describes three measurement schemes that
can be used to recover intracavity absorption information from the
cavity transmitted beams. The two common methods are cavity-enhanced
absorption spectroscopy (CEAS) and cavity ring-down spectroscopy
(CRDS). An intermediate scheme is transient CEAS.

\label{sec:1}
\subsection{Cavity-enhanced absorption spectroscopy}
CEAS refers to measurements of the intracavity absorption that are
made by comparing the power transmitted through the cavity with and
without absorption present.  These measurements can be performed by
either locking the comb to the cavity or by sweeping the comb
frequencies over the cavity modes. For the purposes of this section,
locked measurements will be referred to as steady state CEAS while
measurements involving frequency sweep will be referred to as
transient CEAS. As will be seen in the following section, the signal
to noise ratio achieved in cavity transmission and thus the
absorption sensitivity depends slightly on the type of CEAS that is
implemented.

\indent For steady state CAES measurments, the comb is locked to the
cavity, allowing the intracavity power in each mode to reach steady
state after several cavity lifetimes.  For a single comb component
that is on resonance with a cavity mode, the effect of absorption on
the power of the light transmitted through the cavity is given by
\begin{equation}
\frac{P_{t}(\nu)}{P_{inc}}=\frac{T_1T_2e^{-\alpha (\nu)
L}}{\big|1-\sqrt{R_1 R_2}e^{-\alpha (\nu) L}\big|^2}.
\end{equation}
Here P$_{t}(\nu)$ is the transmitted power, P$_{inc}$ is the power
incident on the cavity, and R$_{1,2}$ and T$_{1,2}$ are the
reflectivity and transmission coefficients of the cavity mirrors at
frequency $\nu$. $L$ is the cavity length and $\alpha(\nu)$ is the
frequency-dependent intracavity absorption \cite{63}\cite{64}. In
the limit that $\alpha (\nu)L \ll 1$ and $R_{1,2} \approx 1$, the
change in transmitted power due to absorption $\alpha (\nu)$ can be
written as
\begin{equation}
\frac{\delta P_{t}(\nu)}{P_{inc}}=\frac{2 F \alpha (\nu) L}{\pi}.
\end{equation}

Equation 5 illustrates that in the limit of low absorption, steady
state CEAS results in an enhancement of the single pass absorption
signal by a factor of $2F/\pi$.  Steady state CEAS has the benefit
that it is the most sensitive of the intracavity absorption
measurement techniques.  However, as discussed in the previous
section, there are several drawbacks to this technique. The first is
the complexity of the comb/cavity servo system since it is necessary
to control both the laser $f_{r}$ and $f_{o}$ to high precision.
Secondly, intracavity dispersion due to the mirrors and sample gas
results in a limited spectral bandwidth over which the comb
frequencies and cavity modes are overlapped. Finally, when locking a
comb to a high finesse cavity, it is difficult to avoid frequency to
amplitude noise conversion, unless a technique similar to NICEOHMS
is employed \cite{65}\cite{66}. Consequently, the sensitivity
benefit of steady state CEAS is often compromised by technical
noise.

\indent Transient CEAS, performed by a frequency sweep between the
comb and cavity modes, overcomes some of the limitations of steady
state CEAS. This scheme is relatively easier to implement due to the
relaxed locking requirements and it solves the problem of limited
spectral bandwidth encountered when locking the comb to the cavity.
The cost of transient CEAS is a reduced sensitivity for absorption
measurements.

To perform transient CEAS, the comb modes are made to overlap the
cavity modes like they would be for steady state CEAS. However,
instead of locking the comb to the cavity, a low modulation
amplitude ($<$10\% $FSR$), fast ($>$1 kHz) sweep is applied to
either the laser $f_{r}$, $f_{o}$, or the cavity $FSR$. For a single
comb component, the effect of intracavity absorption on the
transmitted power is given by
\begin{equation}
\frac{P_{t}(\nu)}{P_{inc}}=T_1T_2\big|\sum_{n=1}^\infty
r_1^{n-1}r_2^{n-1}e^{-(2n-1)\alpha L/2} e^{i\phi _{n} (t)}\big|^2.
\end{equation}
Here $n$ is the number of cavity round trips and r$_{1,2}$ are the
electric field reflection coefficients for the cavity mirrors.  The
phase term is given by
\begin{equation}
\phi _{n} (t)=2\pi(2n-1) (L\frac{\nu (t)}{c}-(n-1)\beta
\frac{L^{2}}{c^{2}}),
\end{equation}
where $\nu (t)$ is the instantaneous frequency of the comb
component, and $\beta$ is its frequency sweep rate
\cite{67}\cite{68}. Besides the lower detection sensitivity,
transient CEAS also has the disadvantage that quantitative
measurements of the intracavity absorption requires careful
calibration. The calibration accuracy is contingent on precise
control of the frequency sweep. Maintaining a constant sweep rate is
important. A common culprit causing non-constant sweep rates is the
hysteresis found in transducers such as piezo electrics.  It is
therefore necessary to either make measurements only in one sweep
direction or determine the actual frequency sweep function and take
measurements when the opposing sweep rates are equal.

\subsection{Cavity ring-down spectroscopy}
Another method for making intracavity absorption measurements is
cavity ringdown spectroscopy (CRDS) \cite{57} \cite{69} \cite{70}
\cite{71}\cite{72}. Light is first injected into the resonant modes
of the optical cavity. When the intracavity power in each cavity
mode reaches a sufficient level, the light incident on the cavity is
switched off. The cavity decay signal is then recorded by a
photodiode with a detection bandwidth exceeding that of the cavity
decay rate. The power in each cavity mode decays exponentially with
a decay constant given by
\begin{equation}
\tau_{cavity}=\frac{1}{FSR(\nu)
Loss}=\frac{1}{FSR(\nu)(1-R_{1}R_{2}e^{-2\alpha(\nu) L})}.
\end{equation}

Depending on the length and finesse of the optical cavity, this
decay time can take on values ranging from tens of nanoseconds to
hundreds of microseconds.  In the limit that $\alpha (\nu) \ll 1$
and $R_{1,2} \approx 1$, the change in cavity decay time due to
absorption $\alpha (\nu)$ can be written as
\begin{equation}
\frac{\delta \tau_{cavity}(\nu)}{\tau_{cavity}}=\frac{F \alpha (\nu)
L}{\pi}.
\end{equation}

For CRDS, the absorption sensitivity in the low absorption limit is
exactly half of the steady state CEAS sensitivity.  This is due to
the fact that for steady state CEAS, the presence of the incident
beam suppresses decay of the field from the input mirror. While in
CRDS, the field decays through both the input and output mirrors. In
the case that the reflectivity of the input and output mirrors are
exactly equal, CEAS is twice as sensitive to intracavity absorption
as CRDS.  In the case of unequal reflectivities, equations 4 and 8
can be used to determine the relative sensitivities.

\indent CRDS has a couple of advantages over the CEAS methods for
measuring intracavity absorption. First, since the laser is
effectively switched off during the measurement time, cavity
ringdown measurements are free of laser intensity noise. Also, since
the finesse of an optical cavity is typically a very stable quantity
that can be measured to high precision, background measurements of
the cavity decay rate without any intracavity absorption can be
performed relatively infrequently. Hence, more time can be devoted
to making absorption measurements rather than calibrating them.  The
only drawback of cavity ringdown spectroscopy is that it requires
fast sampling to record the cavity decay rate. This becomes
challenging in broadband applications where hundreds or even
thousands of channels must be simultaneously sampled to record an
intracavity absorption spectrum.

\subsection{Sensitivity comparison}
\indent To conclude this section we present a comparison of the
sensitivity between each of the three measurement schemes.  This
comparison will be illustrated by the sensitivity of a transient
CEAS measurement as a function of the speed at which a comb
frequency is swept over the cavity mode.  In the limit of an
infinitely slow sweep, the sensitivity of transient CEAS approaches
that of steady state CEAS. In the limit of a very fast sweep speed,
the sensitivity of transient CEAS approaches that of CRDS. Figure 11
shows both the low absorption limit detection sensitivity and the
time-integrated fractional transmitted power in transient CEAS as a
function of the sweeping speed of the incident laser frequency.
\begin{figure}[h]
\resizebox{0.5\textwidth}{!}{%
 \includegraphics{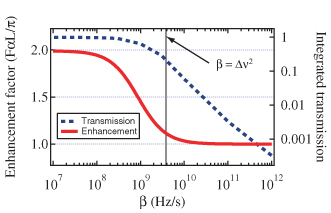} }
 \caption{ Absorption sensitivity and
integrated power of cavity transmission versus sweep speed ($\beta$)
for transient CEAS. Cavity $FSR$ = 400 MHz and mirror reflectivity R
= 0.9995 resulting in a cavity linewidth $\Delta \nu$ = 63.7 kHz. A
time window of 100 $\mu$s was used for integrating the cavity
transmitted signal.}
\end{figure}

For the calculations displayed in Fig. 11, the cavity $FSR$ = 400
MHz and the mirror reflectivity R = 0.9995, resulting in a cavity
lifetime of 2.5 $\mu$s. A time window of 100 $\mu$s was chosen for
integrating the cavity transmitted signal. The center wavelength was
1.525 $\mu$m and the intracavity absorption was 10$^{-8}$ cm$^{-1}$.
At slow sweep speeds, it takes a comb frequency many cavity
lifetimes to scan across the cavity linewidth. Under these
conditions, the intracavity power nearly reaches the steady state
value and cavity transmitted beam power is $P_\alpha=P_o(1-2F\alpha
L/\pi$). Here $P_o$ is the cavity transmitted power with no
intracavity absorption.  For a fast sweep speed, the comb frequency
scans across the cavity resonance in less than a cavity lifetime. In
this case the peak intracavity power is much smaller than the steady
state value, and the majority of the cavity transmitted signal takes
the form of an exponential decay. Since the time integral of an
exponentially decaying signal is proportional to its decay time, the
change in the integrated cavity transmission due to absorption is
just proportional to the change in the cavity ringdown time, thus
linking transient CEAS to CRDS.  When the frequency sweeping speed
($\beta$) is such that the comb frequency traverses the cavity
linewidth in one or a few cavity lifetimes ($\beta \approx \Delta
\nu^2$), changes in the cavity transmission due to intracavity
absorption arise from a combination of steady state CEAS
and CRDS effects.\\
\indent  In practice, it is usually advantageous to sacrifice the
factor of 2 enhancement associated with steady state CEAS and work
in the regime of fast sweeping.  Doing so allows us to avoid
relatively low frequency mechanical noise present in both the laser
and the passive optical cavity. The gain in $S/N$ by making
detection at Fourier frequencies where mechanical noise is low is
typically much greater than the factor of 2 gained via steady state
detection.
\section{Detection schemes}
In the previous section we have outlined a few strategies for making
intracavity absorption measurements. In this section we focus on
broad-bandwidth detection schemes that can be implemented to provide
frequency-resolved and massively parallel detection channels for the
intracavity absorption. Depending on particular applications,
different detection schemes can provide uniquely optimized
performances.

\subsection{Photodiode array or CCD detection}
The simplest way to perform parallel detection of a broadband cavity
transmission signal is to use a dispersive element such as a
diffraction grating followed by imaging optics and a multi-element
photodetector. The detector can take the form of a photodiode array
or a CCD camera. In this configuration, the final detection
resolution is determined by the spectral resolving power of the
grating and the imaging resolution on the detector array. Grating
spectrometers typically provide resolutions on the order of a few to
many tens of gigahertz. The overall spectral bandwidth that can be
recorded by the detector in a single shot is limited by the size of
the detector array and the area over which a flat field of imaging
can be maintained.  Single shot spectral bandwidths range from
hundreds of gigahertz to several terahertz. For a fixed detector
size with $n$ resolvable detection channels and a resolution of
$\delta \nu$, the spectral bandwidth that can be recorded in a
single shot is $n \delta \nu$. This relationship illustrates the
compromise that must be made between single shot bandwidth and
resolution that is encountered when designing a broadband detection
system.

Another issue that arises from multichannel detectors is the rate at
which the array can be read out.  Depending on the detector size and
the type of a multiplexer used for sampling, read-out rates in the
tens of Hz to tens of kHz range can be achieved.  While the later
provides fast frame acquisitions for signal averaging, it is still
too slow to directly observe cavity ringdown decays. Most cavities
require a sampling rate of around 1 MHz to perform accurate ringdown
measurements. To simultaneously sample ringdown signals from a
large-numbered photodiode array, we have developed an electronics
architecture that is capable of sampling and recording ringdown
events from an arbitrarily large photodiode array at rates that can
easily exceed 1 MHz.

A block diagram of the electronics architecture is shown in Fig. 12.
A large photodiode array, consisting of hundreds to thousands of
detector elements, is first divided into sub-arrays of 10 to 50
elements that detect adjacent portions of the optical spectrum. The
photodiode array in Fig. 12 represents one of these sub-arrays.
\begin{figure}[h]
\resizebox{0.5\textwidth}{!}{%
 \includegraphics{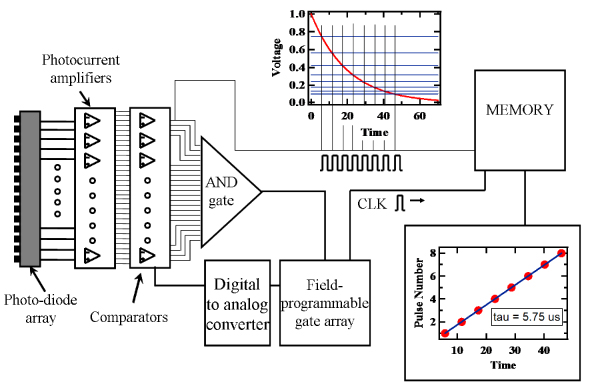} }
\caption{ A circuit for parallel readout of cavity ringdown signals
from a photodiode array.}
\end{figure}
The current from each element in the sub array is converted to a
voltage by a bank of transimpedance amplifiers.  Next, a bank of
comparators is used to compare these photo-voltages against a known
reference voltage that is set by a digital-to-analog converter
(DAC). The reference voltage is chosen to be a fraction of the peak
voltage recorded by one of the detector elements in the sub-array.
As the intracavity field decays, each photo-voltage drops below the
reference voltage and the corresponding comparator emits a pulse
that is recorded by a clocked memory chip. The comparator pulse is
also sent to the input of an AND gate.  When each signal drops below
the DAC voltage, the output of the AND gate switches, instructing
the DAC to set a new reference voltage.  The subsequent reference
voltages are chosen to be a power law series such that an
exponentially decaying ringdown signal will produce a series of
comparator pulses that are equally spaced in time.  This sequence
continues until all signals drop below the lowest reference voltage.
During this process, a field programmable gate array (FPGA) performs
a number of operations to ensure that switching of reference
voltages occurs at the appropriate times and that the ringdown
signal for each channel is recorded correctly.  When the comparator
pulse arrival data is retrieved from the memory chip, a simple
linear fit of the arrival time of the comparator pulses versus pulse
number reveals the ringdown time of that channel.  So far, only a
small prototype of this circuit consisting of four channels has been
built and tested, but the scalability of this architecture will
allow for hundreds to thousands of channels to be simultaneously
sampled with small additional effort.

\subsection{VIPA spectrometer}
One way to overcome the compromise between spectral bandwidth and
resolution is to use a virtually imaged-phased array (VIPA)
spectrometer (Fig. 13) \cite{73}\cite{25}\cite{29}. This
spectrometer uses a tilted solid thin Farby-Perot etalon, oriented
orthogonally to a diffraction grating that is placed after the
etalon, to disperse the cavity transmitted beam into a
two-dimensional pattern, with each spatial element corresponding to
a unique comb frequency.
\begin{figure}[h]
\resizebox{0.5\textwidth}{!}{%
 \includegraphics{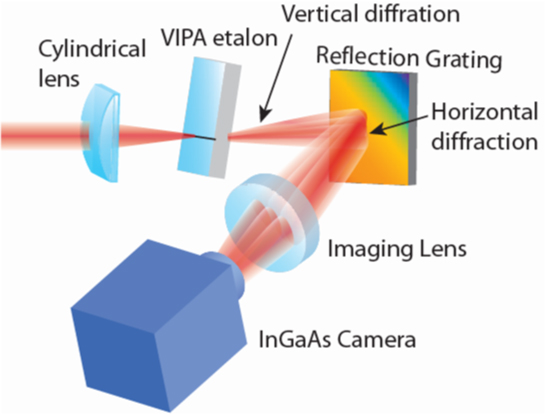} }
\caption{A schematic of the VIPA spectrometer.}
\end{figure}
The 2D spatial pattern of the comb is then imaged onto a CCD camera
to record the frequency-resolved power of the cavity transmission.
The angular dispersion of the VIPA etalon can be 30 to 40 times
greater than the dispersion of an optical grating. As a result,
light dispersed along the tilt of etalon provides a high spectral
resolution, while light dispersed along the orthogonal direction by
the grating acts to resolve the different mode orders transmitted
through the VIPA etalon.

Figure 14 shows a typical VIPA absorption image of the rovibrational
spectrum of CO$_2$ centered at 1.609 $\mu$m.  Each vertical fringe
contains nearly two $FSR$ mode orders of the etalon transmission,
which has been dispersed and resolved in the horizontal direction by
the grating. To resolve the vertical fringes in a VIPA spectrometer,
the grating must have a resolution that is higher than the $FSR$ of
the VIPA etalon.
\begin{figure}[h]
\resizebox{0.5\textwidth}{!}{%
 \includegraphics{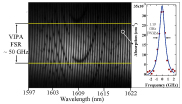} }
\caption{An image of a rotational spectrum of a CO$_2$ vibration
overtone centered at 1.609 $\mu$m recorded with the VIPA
spectrometer. The image contains 25 nm of spectrum with 833 MHz
resolution.  The panel to the right shows the P13 line of the (011)
to (311) vibration overtone band, illustrating the resolution.}
\end{figure}

In Fig. 14, the VIPA etaon has a $FSR$ of 50 GHz while the 1100
lines/mm grating imaged by a $f$ = 20 cm lens is capable of
resolving 25 GHz. The camera recording the 2D pattern has dimensions
256 $\times$ 320 pixels with a pixel pitch of 25 $\mu$m.  The
diffraction limited spot size of the beam imaged on the camera is
~50 $\mu$m such that each resolvable frequency component in the
spectrometer focal plane is sampled by 4 pixels of the camera. The
size of the camera, the grating dispersion and the imaging lens
focal length lead to a 25 nm spectral bandwidth captured in a single
image.  Since there are 120 vertical pixels contained in one $FSR$
of the recorded image, the resulting spectral resolution is roughly
equal to (50 GHz)/[(120 $pixels$)/(2 $pixel$ $resolution$)] = 833
MHz.  To accurately calculate the resolution and wavelength as a
function of the position on the camera requires the generalized
grating equation for the VIPA \cite{74}. In the example presented
here, more than 3500 spectral channels are acquired simultaneously,
far more than that could be achieved in one dimension and with
significantly higher resolution.

\indent  To extract a traditional absorption spectrum from the
recorded VIPA images requires a modest amount of image processing. A
flow diagram of the spectral recovery process is presented in Fig.
15, where a pair of 256 x 320 pixel images are converted into a 25
nm absorption spectrum.  The first step is to locate the nearly
vertical fringes produced within the picture. This is done using the
reference image, which is collected without intracavity absorption.
Fringes are collected by an algorithm that begins at the peak of
each fringe at the middle of the picture (i.e. row 128) and 'walks'
along each fringe up and down to find the peak of each fringe within
the entire picture frame. The intensity of each fringe is converted
into a column vector forming an array that contains frequency
resolved cavity transmitted power information.

\begin{figure}[h]
\resizebox{0.5\textwidth}{!}{%
 \includegraphics{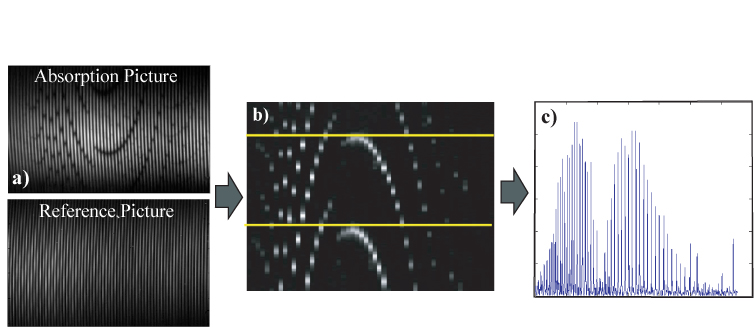} }
\caption{Generation of an absorption spectrum from VIPA images. a)
absorption and reference images are recorded. b) The nearly vertical
fringes in both the reference and absorption images are collected
into column vectors and processed according to $(P_r-P_a)/P_r$ to
form a relative power array.  The relative power array is used to
find the VIPA $FSR$, indicated by the area between the solid lines.
c) The fringe intervals between the solid lines are then collected
and arranged endwise to form a traditional absorption spectrum.}
\end{figure}

\indent To correct for possible drifts in the cavity transmitted
power between the acquisition of the absorption and reference
pictures, an area of low absorption in the absorption fringe array
is equalized in power to the reference fringe array. The two fringe
arrays are subtracted and then divided by the reference fringe
array, resulting in a power-normalized fringe array shown in Fig.
15b.

The $FSR$ interval of the VIPA, shown as the interval between the
two solid horizontal lines in Fig. 15b, is determined from the
recognition of the repetitive patterns of absorption peaks in the
power-normalized fringe array. All unique information about the
molecular absorption is contained within this interval.  Once this
interval is determined, the area of the normalized fringe array
between the horizontal lines is unwrapped into a one dimensional
spectrum of relative power versus wavelength. Here the generalized
grating equation for the VIPA can be used to obtain a high precision
mapping of the position of each wavelength on the normalized fringe
array. Finally, the cavity enhancement equations from Section 5 are
used to convert changes in cavity transmitted power to their
corresponding values of intracavity absorption.

\subsection{Precision scan of $f_r$ and cavity filtration}
Another approach for achieving high spectral resolution over a large
spectral window is cavity-filtered comb detection. Unlike previously
mentioned approaches that use optical dispersers to analyze the
cavity transmission, the cavity-filtered detection uses the
frequency response of the cavity itself to allow only one spectral
component of the frequency comb to be transmitted through the cavity
at a time. In this Section  we present an example of cavity-filtered
detection. The cavity frequency filtration is discussed in Section
4.3. The results presented here are based on a $f_r$ = 99 MHz erbium
fiber mode-locked laser comb, a detection cavity with a $FSR$ of 396
MHz ($FSR$ = 4 $\times f_r$) and a finesse of 6300 at the detection
wavelength 1.525 $\mu$m. The intrinsic cavity linewidth is 63.5~kHz.

\indent To use the cavity-filtration approach for frequency-resolved
detection of cavity transmission, it is necessary to precisely
control the frequencies of both the cavity and comb modes. For the
measurements presented here, the frequency stabilization scheme is
shown in Fig. 16. The cavity is stabilized by locking a particular
cavity resonance (and thus the cavity length) to a
frequency-stabilized Nd:YAG laser at 1064 nm.
\begin{figure}[h]
\resizebox{0.5\textwidth}{!}{%
 \includegraphics{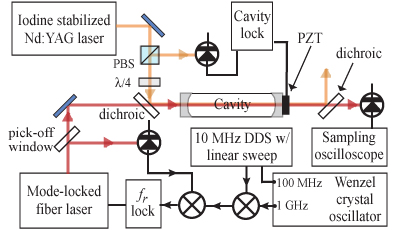} }
\caption{ Schematic for the cavity filtration measurement setup. An
iodine stabilized Nd:YAG laser (1064 nm) is used to stabilize the
cavity frequencies via a piezo electric actuator that controls the
cavity length.  The 10$^{th}$ harmonic of the laser $f_r$ is locked
to a microwave signal that is the mixing product of a 1 GHz Wenzel
oscillator and a 10 MHz DDS.  To scan a spectrum, the DDS frequency
is set such that the laser $f_r$ is detuned from the cavity FSR. The
DDS frequency is then linearly scanned, resulting in a
time-dependent output frequency for the cavity transmission. }
\end{figure}
The cavity length is stabilized to $\lambda$/3200 RMS. For $FSR$
$\sim$400 MHz, this corresponds to stabilization of a cavity
resonance frequency to within 250~kHz. To stabilize the relative
frequency between the cavity and the comb, the tenth harmonic of the
laser $f_r$ is locked to the frequency difference of a 1 GHz
low-phase-noise crystal oscillator and a 10 MHz auxiliary RF signal
from a direct digital synthesizer (DDS).  The time base of the DDS
is slaved to the same low-phase-noise crystal oscillator. The DDS is
used to generate a high precision linear frequency ramp that sweeps
the comb frequencies over the cavity modes.  The DDS makes a 500 Hz
ramp in steps of 23 mHz, each lasting 10.2 $\mu$s for a total
sweeping time of 0.22 s.

\indent To illustrate the utility of this scheme for
frequency-resolved detection of cavity transmission, the cavity was
backfilled with 100 mTorr of C$_2$H$_2$ and 500 Hz scans were taken
at detunings of $\delta f = f_r - FSR$ = 2.5 kHz and 5 kHz. Figure
17 shows the recovered spectral features as a result of these scans.
\begin{figure}[h]
\resizebox{0.5\textwidth}{!}{%
 \includegraphics{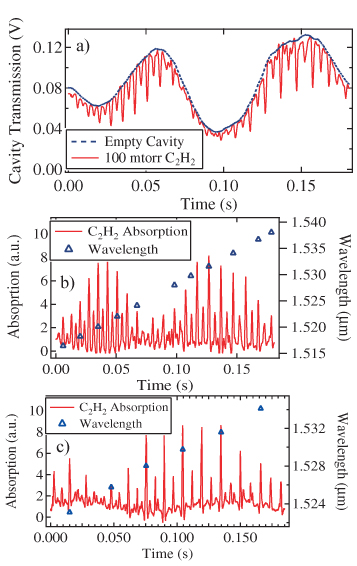} }
\caption{a) The cavity transmission signal with and without 100
mTorr of C$_2$H$_2$ inside the cavity for a 2.5 kHz detuning of the
laser $f_r$ from the cavity $FSR$.  b) The C$_2$H$_2$ absorption
spectrum extracted from a) showing 25 nm of spectrum with 10 GHz
resolution. c) A second sweep where $f_r$ is detuned 5 kHz from the
cavity $FSR$ resulting in a 12.5 nm spectrum with 5 GHz resolution.}
\end{figure}
Since the cavity modes were stabilized to within 250 kHz (effective
$\Delta \nu$), the 2.5 kHz detuned scan resulted in a resolution of
$(\Delta \nu /\delta f)\times f_r$ = 10 GHz.  The available spectral
bandwidth for this scan is $f_r^2/\delta f=3.92$ THz, or 30 nm at
1.525 $\mu$m.  Figure 17a shows 25 nm, or 83\% of the available scan
width, with the C$_2$H$_2$ absorption spectrum clearly extracted in
Fig. 17b.  For the 5 kHz scan, $\delta f$ was increased by a factor
of 2 and consequently the resolution increased and the bandwidth
decreased by the same factor of 2. This leads to a 5 GHz-resolution
spectrum shown in Fig. 17c. The variations in the baselines of these
two scans arise from a combination of cavity/comb coupling noise and
absorptions due to unresolved hot bands of C$_2$H$_2$ that occupy
the same spectral region as the first overtone of the asymmetric C-H
stretch.

A couple of options exist to achieve higher resolution capable of
resolving these lines.  The first and the most intuitive approach is
to increase the detuning $\delta f$ and perform scans of higher
resolution with narrower spectral bandwidth. The second and more
elegant method was recently demonstrated by Gohle et al. \cite{28}.
\begin{figure}[h]
\resizebox{0.5\textwidth}{!}{%
 \includegraphics{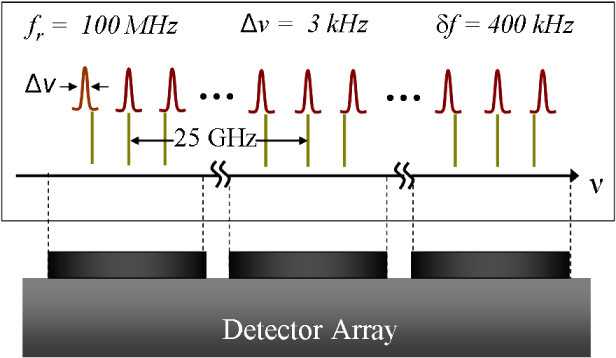} }
\caption{An example of how the comb/cavity detuning ($\delta f$) can
be selected to match the cavity-filtered spectral bandwidth to the
corresponding spatial period of a multichannel spectrometer in
cavity transmission.}
\end{figure}
This method combines cavity filtering with a grating spectrometer in
cavity transmission to achieve high resolution and large spectral
bandwidth in a single system. In a sense, this scheme shares the
same spirit as the VIPA spectrometer, where a high resolution
element (etalon) is combined with a low resolution grating. Whereas
the VIPA scheme uses a second spatial dimension to provide high
resolution, cavity filtration uses temporal scan of the comb
freuqnecies. Fig. 18 shows a schematic of how the cavity filtration
and spatial dispersion can work together to provide high resolution
and large spectral bandwidth. The detuning $\delta f$ is chosen to
match the resolution of the grating spectrometer to the spectral
bandwidth of cavity filtration. In doing so, each resolvable unit of
the spectrometer contains a single comb component at each step
during the frequency scan. For the example shown in Fig. 18, each
scan covers 250 cavity modes requiring a scan time of at least 250
cavity lifetimes for single comb mode resolution.  For typical
cavities, 250 cavity lifetimes corresponds to a scan time of one to
tens of milliseconds.

\section{Applications}
The final Section presents a series of CE-DFCS measurements,
beginning with initial demonstrations of these techniques followed
by more advanced and recent measurements.  This overview is meant to
highlight the general capabilities of these systems as well as the
recent technological advances that have allowed significant
simplifications of experimental setups. The measurements presented
here include trace detection of room temperature gas samples, breath
analysis, and detection and characterization of cold molecules in
supersonic jets.

\subsection{Initial demonstrations}
Gherman and Romanini performed broadband CEAS using a Ti:sapphire
mode-locked laser to probe the high overtone spectrum of C$_2$H$_2$
at 860nm \cite{26}. This system performed transient CEAS
measurements using a cavity with a finesse of F = 420 and achieved
an absorption sensitivity of $2\times 10^{-7}$ cm$^{-1}$
Hz$^{-1/2}$.  A high resolution grating spectrometer and a CCD
camera were implemented in cavity transmission to provide 14 nm of
single shot bandwidth with a resolution of 6 GHz.

\indent Thorpe \emph{et al.} demonstrated frequency-comb-based CRDS
using a Ti:Sapphire laser-based frequency comb at 800 nm to probe
the high overtone spectrum of C$_2$H$_2$, H$_2$O, NH$_3$, and a week
forbidden electronic transition in O$_2$ \cite{27}.  An example of
these measurements is provided in Fig. 19. This system used a
$\mathcal{F}$ = 3300 cavity and cavity ringdown detection to achieve
an integrated absorption sensitivity of 10$^{-8}$ Hz$^{-1/2}$
(2.5$\times10^{-10}$ cm$^{-1}$Hz$^{-1/2}$).
\begin{figure}[h]
\resizebox{0.45\textwidth}{!}{%
 \includegraphics{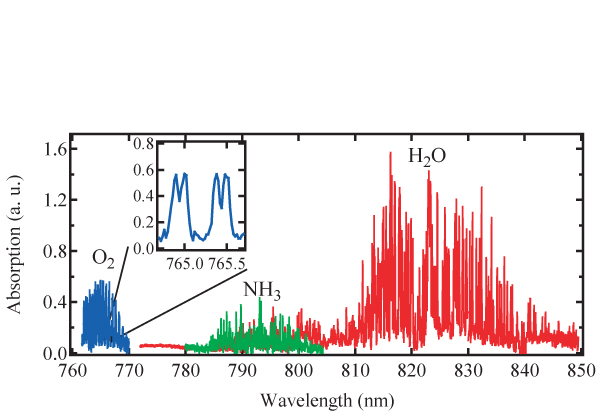}
} \caption{The spectra of H$_2$O, NH$_3$, and O$_2$ from 760 nm to
850 nm.}
\end{figure}
In cavity transmission a grating spectrometer and CCD were
implemented to record 20 nm of single shot spectra with 25 GHz
resolution. To achieve adequate time resolution of the ringdown
signals, a rotating mirror was installed in the grating spectrometer
such that time decay of the cavity transmitted power was streaked
across the vertical columns of the CCD \cite{75}.  Using this
technique the cavity decay signal recorded by the CCD exhibits
decaying power as a function of vertical position on the detector.
The decay versus position signal is converted to decay versus time
using the mirror rotation rate and the distance between the mirror
and CCD.  Finally, the decay signal is fit to extract the cavity
ringdown time.  Each column of the CCD simultaneously records a
decay signal at a different wavelength, resulting in a single-shot,
broad bandwidth ringdown spectrum.

While these initial systems provide good sensitivities for the study
of high overtone spectra, the weak transition strengths of high
overtones make them not very desirable for applications such as
trace detection. Also, the resolution of the initial systems was too
low to resolve the Doppler linewidths of room temperature gases and
therefore limited the sensitivity of molecular concentration
measurements.  In response to these limitations, recent work of
Thorpe \textit{et al.} at 1.5 $\mu$m with Er$^+3$ fiber laser-based
frequency combs aided by high resolution spectrometers in cavity
transmission has drastically increased the detection sensitivity and
spectral resolution available with CE-DFCS systems \cite{29}.

\subsection{Trace detection at 1.5 $\mu$m}
The trace gas measurements presented here were made with an
Er$^{+3}$ fiber comb with $f_r$= 99 MHz.  The laser provided a 100
nm spectrum from 1.5 to 1.6 $\mu$m which could be shifted to 1.6 -
1.7 $\mu$m using a Raman shifting amplifier \cite{76}. The comb was
coupled to a $FSR$ = 99 MHz and $\mathcal{F}$ = 30000 cavity via
dithering and transient CEAS measurements were performed. A VIPA
spectrometer was used in cavity transmission to achieve a 25-nm
single shot bandwidth and 800 MHz resolution.

\indent For these measurements, there are several effects to
consider in order to optimize the spectrometer performance.  The
equation that relates the absorption coefficient measured by the
spectrometer to the gas concentration is
\(\alpha(\nu)=nS_{i}g(\nu)\).  Here, \(\alpha\) is the absorption
per unit length (cm$^{-1}$), $n$ is the gas concentration
(cm$^{-3}$), $S$\(_{i}\) is the line intensity (cm), and $g$($\nu$)
is the lineshape (cm). The values for $S$\(_{i}\) can be measured,
found, or inferred from a variety of references \cite{77}\cite{78}.
However, the lineshape depends on the environmental conditions of
the measurement.  In particular, choosing the appropriate
intra-cavity gas pressure is important for maximizing the molecular
absorption while not diminishing spectral resolution via collision
broadening.

The gases that were measured and analyzed in our 1.5 $\mu$m CE-DFCS
system include: CO, CO$_{2}$, C$_{2}$H$_{2}$, C$_{2}$H$_{4}$,
CH\(_{4}\), and NH\(_{3}\).  Parts of the recorded spectra for each
of these gases are shown in Fig. 20 and Fig. 21.
\begin{figure}
\resizebox{0.45\textwidth}{!}{%
 \includegraphics{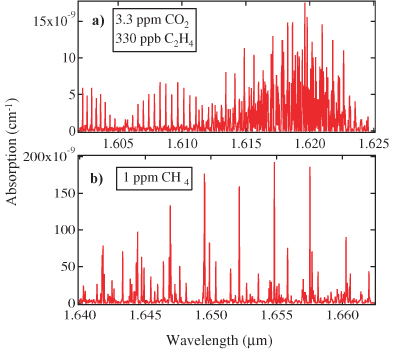} }
\caption{The rovibrational overtone spectra of a) 330 parts per
billion (ppb) C$_2$H$_4$ and 3.3 ppm CO$_2$ and b) 1 ppm CH$_4$.}

\end{figure}
Since these trace gases are measured in a host gas of primarily
N$_{2}$, the air broadening coefficient adequately describes the
observed collision broadening. For measurements presented in this
Section, an intracavity gas pressure of 250 Torr was chosen such
that the average linewidth for a gas under measurement was 1.8 GHz,
resolvable by the 800 MHz resolution of our system. \\
\begin{figure}[h]
\resizebox{0.45\textwidth}{!}{%
 \includegraphics{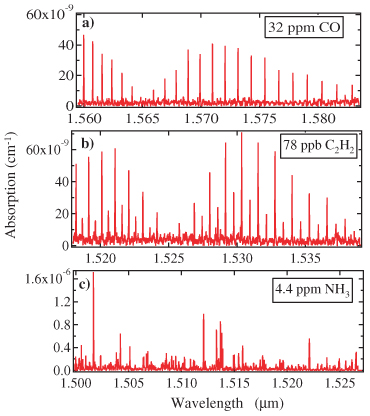}
} \caption{The rovibrational overtone spectra of a) 32 ppm CO, b) 78
ppb C$_2$H$_2$, and c) 32 ppm CO, and c) 4.4 ppm NH$_3$.}
\end{figure}
\indent For the molecules listed above, the spectral window between
1.5 $\mu$m - 1.7 $\mu$m contains more than 10,000 individual lines
for measurements at 296 K.  To handle this large amount of
information, we created a database containing the center frequency,
line intensity, and the air-broadening coefficient of more than 60\%
of the existing lines. This database was compiled from a variety of
sources \cite{77}\cite{78}\cite{6}.  Using this database, gas
concentrations were determined for lines of interest from a recorded
spectrum by performing a modified Voigt fit that also included the
resolution of the VIPA spectrometer.  By finding the value of the
instrument resolution that provided the best fit to the
experimentally recorded lineshapes, an instrument resolution of 800
MHz was experimentally verified.

\indent Calibration of the spectrometer was performed using three
gas mixtures containing a total of four target gases consisting of
10 parts per million (ppm) CO\(_{2}\), 1 ppm CH\(_{4}\), 1 ppm
C\(_{2}\)H\(_{4}\), and 1 ppm NH\(_{3}\), all mixed with balanced
N\(_{2}\).  The accuracy of trace gas concentrations was specified
to within 1\% of the total concentration by a commercial supplier of
standard gas. Since the spectral lines from these four
\begin{table}
\caption{Spectral location and minimum detectable concentration
(MDC) of measured molecules using the 1.5 $\mu$m CE-DFCS.}

\begin{tabular}{lll}
\hline\noalign{\smallskip}
Molecule & Wavelength ($\mu$m) & MDC  \\
\noalign{\smallskip}\hline\noalign{\smallskip}
CO & 1.56-1.62 & 900 ppb \\
CO$_2$ & 1.56-1.62 & 770 ppb \\
CH$_4$ & 1.63-1.69 & 10 ppb \\
C$_2$H$_2$ & 1.51-1.54 & 3 ppb \\
C$_2$H$_4$ & 1.61-1.65 & 33 ppb \\
NH$_3$ & 1.50-1.55 & 15 ppb \\
\noalign{\smallskip}\hline
\end{tabular}
\end{table}
molecules cover the entire spectral region from 1.5 \(\mu\)m to 1.7
\(\mu\)m, we were able to verify the calibration of the absorption
loss provided by the cavity transmission equation (eq. 6) for the
entire spectral region. Finally, we made mixtures of CO and
C\(_{2}\)H\(_{2}\) by mixing pure samples of these gases with
N\(_{2}\) to determine the minimum detectable concentrations of
these gases. These results, along with their detection wavelengths,
are listed in Table 1.

\subsection{Isotope ratio measurements}

A powerful demonstration of the sensitivity, dynamic range, and
resolution of our spectrometer can be made via isotope measurements.
During the course of measuring the spectra of CO, CO$_2$,
C$_{2}$H$_{2}$, H$_2$O, and CH$_4$, we recorded spectral regions
that contain multiple isotopes of all of these molecules. Figure 22
shows the spectra of CO and C$_2$H$_2$ containing multiple isotopes.
\begin{figure}[h]
\resizebox{0.45\textwidth}{!}{%
 \includegraphics{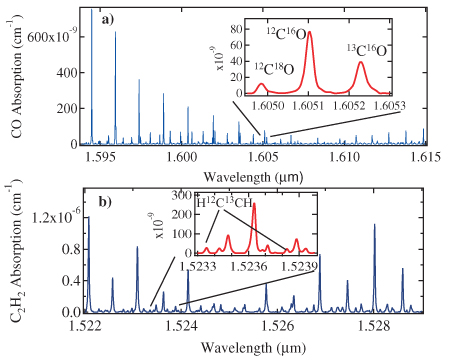}
} \caption{a) The isotope spectrum of CO near 1.605 $\mu$m at a CO
concentration of 4.5 parts  per thousand. b) The isotope spectrum of
C$_2$H$_2$ near 1.524 $\mu$m at a C$_2$H$_2$ concentration of 2.2
ppm.  The unlabeled lines in the C$_2$H$_2$ spectrum are
$^{12}$C$_2$H$_2$ while the labeled lines are H$^{12}$C$^{13}$CH.}
\end{figure}
For both CO and C$_{2}$H$_{2}$, the relative abundances of the
isotopes measured by our system matched those found in the HITRAN
database to within 0.4\%. We also estimated the minimum
concentration of CO and C$_{2}$H$_{2}$ that would allow detection of
the second most abundant isotope. These concentrations were 200 ppm
for CO and 200 ppb for C$_{2}$H$_{2}$.

\indent  The ability to accurately measure molecular isotope ratios
has found important applications in many fields of science.  For
instance, doctors use stable isotope ratios of $^{13}$CO$_2$ on the
breath of patients to detect \emph{Heliobacter Pylori} for
diagnosing ulcers \cite{79}. Geologists use stable isotopes of
H$_2$O to investigate the transport of groundwater sources
\cite{80}. Climatologists and paleontologists use stable isotopes of
O$_2$ to determine the past temperature of the Earth's atmosphere
from ice cores \cite{81}. These examples represent just a few of the
useful applications of stable isotope ratio measurements.

\subsection{Breath analysis}

One emerging application of trace gas detection is human breath
analysis.  Human breath contains trace quantities of more than 1000
molecules \cite{82}.  Abnormal concentrations for many of these
different molecules, called 'biomarkers', have already been
correlated to specific diseases and health conditions \cite{83}.
CE-DFCS is particularly well suited for breath analysis because of
its wide spectral coverage allowing for detection of many different
biomarkers simultaneously.  In the 1.5 - 1.7 $\mu$m region, there
are more than 10 biomarkers that exist in human breath in detectable
quantities. Figs. 23 and 24 show two sets of measurements that
detect a couple of these biomarkers in human breath samples. These
measurements were performed with the 1.5 $\mu$m system described in
the previous section.  Detailed descriptions of these measurements
can be found in \cite{29}.

The first set of breath measurement, shown in Fig. 23, records
absorptions of three different isotopes of CO$_2$. This measurement
is of a medical interest because the ratio of $^{13}$C$^{16}$O$_2$
to $^{12}$C$^{16}$O$_2$ can be used to determine if a patient is
infected with \emph{Heliobacter pylori}, a common cause of ulcers.
This breath test is conducted by having the patient ingest a
$^{13}$C-labeled urea and then measuring the ratio of
$^{13}$C$^{16}$O$_2$ to $^{12}$C$^{16}$O$_2$ on the patients breath
\cite{84}.  Increased levels of $^{13}$C$^{16}$O$_2$ indicate the
presence of \emph{H. pylori}, a bacteria that helps to convert the
$^{13}$C-labeled urea into $^{13}$CO$_2$ before being exhaled in the
patients breath.
\begin{figure}[h]
\resizebox{0.45\textwidth}{!}{%
 \includegraphics{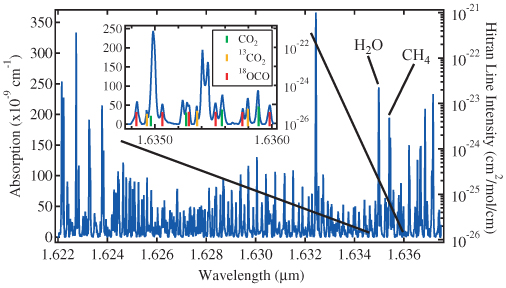}
} \caption{a)Breath spectrum of CO$_2$ isotopes near 1.63 $\mu$m.
The inset shows a zoomed-in portion of the CO$_2$ landscape. The
spectral region between 1.622-1.638 $\mu$m contains 78 CO$_2$ lines,
29 $^{13}$CO$_2$ lines, and 62 $^{18}$O$^{12}$C$^{16}$O lines making
it an ideal spectral region for determining isotope ratios.}
\end{figure}
In the spectral region shown in Fig. 23, there are 78 CO$_2$
absorption lines and 29 $^{13}$CO$_2$ lines.  The absorption due to
each of these lines is used to determine the final
$^{13}$C$^{16}$O$_2$ to $^{12}$C$^{16}$O$_2$ ratio which achieves an
accuracy of 0.4\%.

The second breath measurement shown in Fig. 24 illustrates how
environmental and behavioral factors that determine health can be
monitored via breath analysis.  Figure 24 shows the breath spectra
of two students.  Student 1 is a smoker who had a cigarette 15
minutes prior to providing the breath sample. Student 2 is a
non-smoker who has relatively low exposure to carbon monoxide.
Figure 24 shows that the breath of student 1 has a
\begin{figure}[h]
\resizebox{0.45\textwidth}{!}{%
 \includegraphics{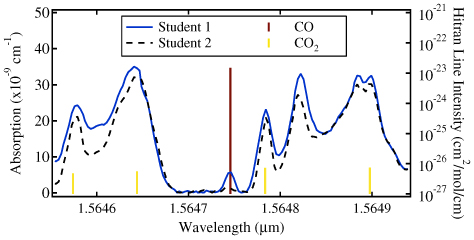}
} \caption{a)Breath spectra near 1.565 $\mu$m showing the difference
between CO concentration in the breath of student 1 (smoker) versus
student 2 (non-smoker). The vertical bars indicate the wavelengths
and line intensities of CO$_2$ and CO lines according to the HITRAN
database.}
\end{figure}
high CO concentration (6 ppm), student 2 has a low CO concentration
($<$ 1 ppm).  While smoking provides an example of the behaviorial
effect on the CO level in breath, environmental conditions can also
lead to elevated CO breath levels. For instance, individuals who
spend a large portion of their day in congested traffic or people
live with elevated levels of CO at home are also prone to elevated
CO levels in their breath \cite{85}.

The breath measurements presented here provide a modest glimpse of
the health information that can be acquired by analyzing a patients
breath.  In the 1.5 $\mu$m spectral region alone there are also
strong absorptions from ammonia which can be used to detect kidney
failure \cite{86}, acetone which is an indicator for diabetes
\cite{87}, and ethane and ethylene for detecting lung cancer
\cite{2}.

\subsection{Spectroscopy of cold molecules}

So far, measurements presented in this work have been on room
temperature gas samples at relatively high pressures and therefore
it is not required to employ higher spectral resolution than what is
provided by the VIPA spectrometer. In this Section, we discuss a
technique that uses precise control of the comb and cavity
frequencies in conjunction with a VIPA spectrometer to extend the
spectral resolution far beyond what is provided by the VIPA alone.
We present some measurements on a cold molecular beam of 10\%
C\(_{2}\)H\(_{2}\) in a supersonic expansion of argon carrier gas
that demonstrate these extended capabilities.

The first step to performing these measurements is to stabilize
either the cavity or comb modes.  In the measurements presented
here, the cavity modes are stabilized as shown in Fig. 25. In this
system, the frequency comb \(f_r = 99\) MHz and the cavity $FSR$ =
396 MHz.
\begin{figure}[h]
\resizebox{0.5\textwidth}{!}{%
 \includegraphics{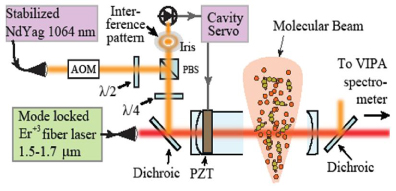}
} \caption{Schematic of the cold molecular beam detection system. An
iodine stabilized Nd:YAG laser stabilizes the cavity mode
frequencies to $\Delta \nu$ = 250 kHz.  An Er$^{+3}$ fiber comb is
coupled into a $\mathcal{F}$ = 6380 and FSR = 400 MHz cavity that is
oriented perpendicular to a pulsed supersonic beam. The cavity
transmitted beam is dispersed and detected using a VIPA
spectrometer.}
\end{figure}
This mismatch of $f_r$ and $FSR$ has the unfortunate side effect
that only 25$\%$ of the comb components are coupled into the cavity
for spectroscopy.  Ideally, a laser with a high $f_r$ (400 MHz $\le
f_r \le$ 800 MHz) would be coupled to a cavity of the equivalent
$FSR$. Since a high-$f_r$ laser was not available, we used a
high-$FSR$ cavity to make the frequency spacing between the comb
modes in cavity transmission close to the resolution of the VIPA
spectrometer.  This configuration allows high resolution features to
be retrieved with high measurement contrast.
\begin{figure}[h]
\resizebox{0.45\textwidth}{!}{%
 \includegraphics{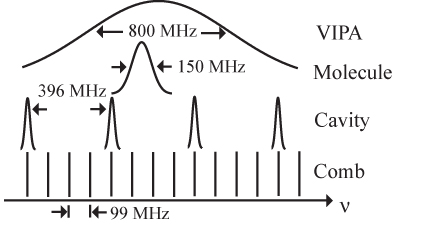}
} \caption{High resolution can be achieved using a combination of
VIPA resolution and frequency stabilized cavity modes.}
\end{figure}

The diagram in Fig. 26 provides a visual description of how higher
resolution is achieved even if the cavity $FSR$ is not greater than
the VIPA resolution. When one of the cavity modes within the
resolution envelope of the VIPA experiences loss due to absorption,
the VIPA channel registers a loss that is diluted by the number of
non-interacting modes contained within the VIPA resolution envelope.
If only a single mode lies within each VIPA pixel, then the
absorption experienced by this single mode can be extracted, leading
to an enhanced signal contrast and improved spectral resolution. In
the case of measurements presented here, $\sim$4 comb modes are
present under each VIPA resolution envelope.  The effect of multiple
comb modes under each VIPA resolution channel is to dilute the
absorption signal recorded by that channel.  The relatively large
number of comb modes per VIPA resolution channel is due to the
Lorentzian transfer function of the VIPA.  By stabilizing the comb
with respect to the cavity, precise frequency sweep of the comb and
cavity modes can then be executed to ensure that only one comb mode
is coupled out at a time under the VIPA resolution. A
high-resolution absorption landscape can then be mapped out with the
resolution limited only by the linewidth of a single comb component.

\indent For the cold C\(_{2}\)H\(_{2}\) measurements presented here,
the cavity was stabilized using an iodine-stabilized Nd:YAG laser at
1.064 $\mu$m.  The cavity length was locked to within
$\lambda$/3200, resulting in a frequency stability of each cavity
mode of 250 kHz. The cavity modes were then scanned using an
acousto-optic modulator (AOM) that varied the frequency of the
Nd:YAG laser. To scan a complete $FSR$ at 1.5 $\mu$m, the frequency
of the 1.064 $\mu$ laser must be scanned $1.5/1.064 \times 400$ MHz
$\approx$ 600 MHz. Figure 27 shows a full $FSR$ scan of the cavity
modes at the peak absorption wavelengths of eight of the lowest
lying rotational lines in the $\nu_1+\nu_3$ C$_{2}$H$_{2}$ overtone
spectrum.  Due to the broadband nature of the VIPA spectrometer,
these traces (plus thousands of others) were recorded during a
single frequency scan by the AOM.

\begin{figure}[h]
\resizebox{0.45\textwidth}{!}{%
 \includegraphics{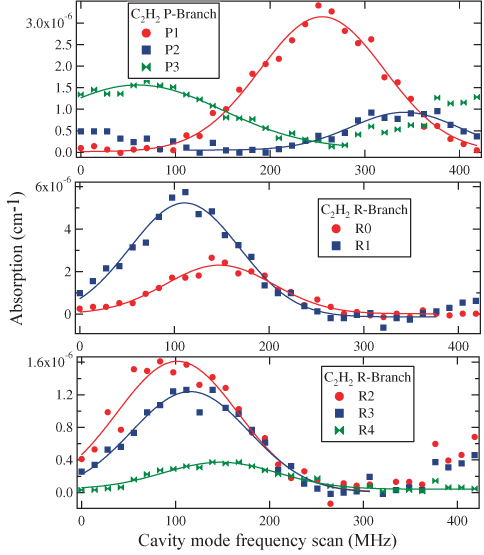}
} \caption{Precise scans of the cavity mode frequencies over one
cavity FSR (400 MHz), revealing spectroscopic features with
linewidths below the VIPA and the $FSR$ resolution limits. Panels
(a), (b), and (c) display the lowest R- and P-branch rotational
lines of the C$_2$H$_2$ $\nu_1+\nu_3$ overtone band detected in a
supersonic cold molecular beam, yielding a complete characterization
of the velocity, internal energy, and density distributions of the
beam. }
\end{figure}

The information recovered from scans such as those shown in Fig. 27
is substantial.  Due to the high frequency resolution of this
technique, accurate lineshapes of the molecular absorption can be
measured to extract both velocity and collision rate information for
the beam.  The broadband nature of this method allows for
simultaneous acquisition of absorption information for many
rovibrational lines, yielding rapid temperature measurements for the
internal degrees of freedom of the molecule. Finally, because the
absorption loss is normalized to the mirror loss, the density of the
molecular beam can be characterized to high precision. Combining
these capabilities while making measurements across different beam
positions allows complete characterization of a cold molecular beam.

\section{Conclusion}
The measurements and techniques presented in this review represent
the current state of the art for CE-DFCS.  As this is a relatively
new field, the capabilities of these systems and the techniques
available are improving rapidly.  Already CE-DFCS has demonstrated
detection of molecular concentrations in the low ppb range for many
molecules with spectral resolutions as low as 250 kHz.  The
remarkable aspect about CE-DFCS is that this high sensitivity and
resolution can be implemented across large spectral bandwidths,
realizing a true real-time scenario for rapid detection.

As applications of these systems become more demanding, the current
limits of CE-DFCS systems will certainly be overcome. Directions for
improving CE-DFCS performance include new broadband cavity designs
such as prism cavities mentioned in Section 3.3. Such cavities would
greatly increase the number of molecules that can be probed with a
single CE-DFCS system. Also, extending precise frequency comb
sources to the mid-IR spectral region, with a reasonable useful
power, will allow much enhanced absorption sensitivity by detection
of strong fundamental vibration transitions, reducing the minimum
detectable concentration of many molecules to the parts per trillion
level.

Aside from trace detection, new applications for CE-DFCS include
detection and characterization of cold molecules, including their
dynamics and interactions. High-resolution coherent control of
atomic and molecular systems similar to work already performed using
free-space DFCS \cite{19}\cite{20} represents another exciting
possibility.

We thank many of our colleagues who have contributed to the work
presented here. In particular, A. Marian and M. Stowe were involved
in the initial demonstration of direct frequency comb spectroscopy.
R. J. Jones and K. Moll helped develop the cavity-enhanced frequency
comb spectroscopy. D. Balslev-Clausen and M. Kirchner participated
in the demonstration of breath analysis via CE-DFCS. F. Adler and K.
Cossel are helping further expand the capabilities and various
applications of CE-DFCS. We also thank T. Schibli, A. Pe'er, D.
Yost, S. Diddams, and D. Hudson for many useful discussions and
technical help. Funding support is provided by AFOSR, DARPA, NIST,
and NSF.


\begin{thebibliography}{}
\bibitem{61}The work at JILA has greatly benefited from collborative coating runs from R.  Lalezari of
Advanced Thin Films. Mention of company or product name is for the
purpose of technical communications and does not represent an
endorsement of NIST.
\end{thebibliography}

\end{document}